\documentclass{aa}
\usepackage{graphicx}
\usepackage{txfonts}
\usepackage{natbib}
\usepackage{comment}
\usepackage{textcomp}
\usepackage{gensymb}
\usepackage{bm}

\usepackage{color}
\definecolor{red}{rgb}{1,0,0}
\definecolor{myblue}{rgb}{0, 0, 1}
 \bibpunct{(}{)}{;}{a}{}{,} 
\begin{document} 
\title{Detection of a radio-filled X-ray cavity within the interstellar medium of NGC 5141}
\subtitle{}
\titlerunning{X-ray cavity within NGC 5141}
\author{Duccio Macconi\inst{1,2} \thanks{E-mail: duccio.macconi2@unibo.it}, Paola Grandi\inst{2}, Myriam Gitti\inst{1,3}, Cristian Vignali\inst{1,2}, Eleonora Torresi\inst{2}, Fabrizio Brighenti\inst{1,4}}
\authorrunning{Macconi D. et al.}
\institute{Dipartimento di Fisica e Astronomia, Alma Mater Studiorum, Universit\`a degli Studi di Bologna, Via Gobetti 93/2, I-40129 Bologna, Italy \and
INAF-Osservatorio di Astrofisica e Scienza dello Spazio di Bologna (OAS), Area della Ricerca CNR, Via Gobetti 101, I-40129 Bologna, Italy \and
INAF, Istituto di Radioastronomia di Bologna (IRA), Via Gobetti 101, I-40129 Bologna, Italy \and
University of California Observatories/Lick Observatory, Department of Astronomy and Astrophysics, University of California, Santa
Cruz, CA 95064, USA}

\abstract
{We present the first \textit{Chandra} detection of a single X-ray cavity within the interstellar medium of the small Fanaroff-Riley type I (FRI) radio galaxy NGC 5141. The X-ray surface brightness depression, located  $\approx 4$ kpc away from the galaxy center, is projected on the northern radio lobe, which is completely contained within the galaxy.
The thermal gas surrounding the cavity, which extends to $\simeq$ 20 kpc, has a bolometric X-ray luminosity (0.1 - 100 keV) of L${_X}\approx2\times10^{40}$ erg s$^{-1}$ and a temperature of $kT\approx0.8$ keV.
We calculated the total energy (E$_{cav} = 4PV \approx 10^{55}$ erg) required to inflate the cavity and its age ($t_{cav}\approx 9$ Myrs),  assuming that it is filled with relativistic particles and rises buoyantly. The inferred total cavity power is as low as P$_{cav}=E_{cav}/t_{cav}\approx6\times10^{40}$ erg s$^{-1}$, which is the lowest one among the radio-filled systems.
Comparing $P_{cav}$ to the bolometric X-ray luminosity (i.e., the cooling luminosity), we conclude that NGC 5141's central active galactic nucleus can heat the interstellar medium and balance its cooling luminosity, confirming that the  $P_{cav}-L_{cool}$ relation, mainly tested on groups and clusters, also works for such a low-power system.}

 \maketitle
\section{Introduction}
\label{introduction}
Over the years, it has become clear that active galactic nuclei (AGN) coevolve with their host galaxies \citep[see e.g.,][]{Magorrian1998,Gebhardt2000,Ferrarese2000,Tremaine2002,Greene2006}. 
The energy emitted by an AGN can heat up and/or remove the cold gas reservoir present in the host galaxy's interstellar medium (ISM), and therefore effectively halt the formation of new stars.
Two kinds of AGN feedback are known: the radiative one, which occurs when the central black hole accretes near the Eddington limit (also known as quasar or wind mode), and the radio mode, which occurs when the AGN has powerful jets interacting with a hot dense halo (also named the kinetic mode, see, e.g., \citealt{fabian2012} for a review). \\
The feedback from AGN is also invoked as a natural solution to the cooling problem in clusters \citep[see e.g.,][]{mcnamara2007,McNamara2012,fabian2012}. The classical cooling flow model \citep{fabian1994} predicted cooling times of the gas in the cluster core shorter than 1 Gyr, implying strong accretion onto the central galaxies (from hundreds to thousands of solar masses per year) and large star formation \citep{birzan2004}. This was not confirmed by the \textit{Chandra} and XMM-\textit{Newton} observations \citep[see, e.g.,][]{Makishima2001,tamura2001}; X-ray studies showed that in many cases, AGN are energetically able to  balance radiative losses from the intracluster medium (ICM) \citep{rafferty2006} and that a significant fraction of the  energy of the radio lobes, powered by AGN, is dissipated within the cooling flow region \citep{Churazov2002}. These results indicated that the dissipation of energy, propagating through the ICM from a central radio source, can balance radiative cooling \citep{Peterson2006}. The radio mode AGN feedback signature consists of a surface brightness depression in the hot surrounding gas (ICM; intragroup medium, IGM; or ISM), excavated by the jet during its propagation. These cavities constitute a fundamental (and direct) tool to estimate the jet's kinetic power and explore the feedback, simply comparing the jet heating with the cooling of the surrounding hot gas. At first, X-ray cavities were found in clusters \citep[e.g.,][]{Boehringer1993,Carilli1994}, then also in small groups and isolated galaxies. However, clusters remain the environment where the vast majority of X-ray cavities are found by far because of their high X-ray surface brightness. Conversely, X-ray cavities in isolated galaxies are the most elusive and they are essentially confined to giant elliptical galaxies \citep[see e.g.,][]{birzan2004,nulsen2007,Cavagnolo2010,birzan2020}.
Here we report the robust case of a non-giant galaxy (NGC 5141) hosting an X-ray cavity in its hot atmosphere.\\
The X-ray cavity detection within NGC 5141 is particularly interesting because of the following: (i) NGC 5141 hosts a known small radio galaxy, whose emission is spatially contained within the optical galaxy; (ii) it enables us to test the heating versus cooling AGN feedback scaling relation to the lowest end of radio galaxies power and size; (iii) the galaxy has a lenticular or elliptical uncertain classification and (to the best of our knowledge) would be the third S0 with detected cavities in  the literature after NGC 4477 (\citealt{Li2018}) and NGC 193 (\citealt{bogdan2014}); and (iv) NGC 5141 shows signs of an interaction with a galaxy companion $\approx50$ kpc away.\\
Throughout the paper the  cosmological parameters adopted are as follows: $H_0=70$ km s$^{-1}$ Mpc$^{-1}$, $\Lambda_\Omega=0.73$, and $\Lambda_m=0.27$ \citep[in agreement with the {\it Planck 2015 Results}, see][]{PlanckCollaboration2016}.
With the adopted cosmology, 1" $\approx0.353$ kpc at the redshift of the galaxy (z=0.01738\footnote{From NED \url{https://ned.ipac.caltech.edu/}}).

\section{The source}
\label{sec_source}

NGC 5141 (a.k.a. UGC 08433) is a local source (z=0.01738), which is part of a small group of six galaxies \citep{ramella1989}. It forms a dual galaxy system with  NGC 5142, which is located $2.3'$ to the northeast ($\approx50$ kpc, see Figure \ref{img-x-feat}-\textit{left panel}; \citealt{soares1989}). 
It is a clear example of an interacting galaxy, as suggested by its nonelliptical isophotes \citep{Gonzalez-Serrano1993}. Its disturbed morphology is probably the cause of a controversial classification.
It was classified as lenticular \citep{nilson1973,Condon1988}, spiral-barred \citep{Courteau2003}, and as elliptical galaxy \citep{Calvani1989,Veron-Cetty2001,Pagotto2017}. \cite{Willett2013} classified the source as a smooth round galaxy with no disk and peculiar features. \cite{Huertas-Company2011}, who performed a Bayesian automated classification of galaxies within the Sloan Digital Sky Survey (SDSS DR7), quantified the classification as uncertain, indicating a probability of $65 \%$ and of $32\%$ that NGC 5141 is an elliptical and an S0, respectively.
Signs of an interaction between NGC 5141 and its companion were also provided by \cite{Emonts2010}.
They found two clouds of resolved HI emission of $\sim6.9\times10^7$ M$_{\odot}$ at a distance of 20 kpc from the center of NGC 5141 in the companion direction  (NGC 5142), interpreted as a bridge-tail structure. HI absorption was also observed on the radio continuum of the radio galaxy. The absorbing gas is slightly extended in the same direction as the emission clouds, and it is probably part of the same large-scale gas structure. NGC 5141 was studied by \cite{ocana2010} for its molecular gas emission lines (CO): they calculated an upper limit on the molecular gas within NGC 5141 of $M_{H_2}<5\times10^8$ M$_{\odot}$. 

Being part of the GALEX-SDSS-WISE legacy, the host properties of NGC 5141 are well studied \citep{Salim2016,Salim2018}. 
Classified as a low excitation radio galaxy (hereafter LERG) by \cite{best2012} based on its narrow-line region emission line ratios, NGC 5141 hides an inefficient accretion flow powered by a black hole mass of M$_{BH}=(4\pm2)\times10^8$ M$_{\odot}$, as estimated from its H-band magnitude, following the correlation by \cite{Marconi2003}.

NGC 5141 has a stellar mass of $log\frac{M}{M_{\odot}}=10.97\pm0.09$ M$_{\odot}$ and a star formation rate of $log(SFR)=-0.529\pm0.054$ M$_{\odot}$ yr$^{-1}$, which are both typical of an early-type galaxy \citep{Salim2016,Salim2018}. The age of NGC 5141 is $t_{sys}\geq 12$ Gyrs as calculated by \cite{Maraston2005} and \cite{Maraston2009}\footnote{They fit the SDSS DR8 galaxies with stellar  spectral templates, considering two stellar initial mass functions (IMFs, Salpeter and Kroupa models) and two stellar population (SP) templates (passive and star-forming galaxies). In this framework, NGC 5141 has a best-fit age ranging from 11.75 Gyrs to >13 Gyrs, taking different IMF and SP.}.
In the central region, an Hubble space telescope (HST) study indicates the presence of a central dust lane with a size of  $\approx$2" along the galactic plane \citep{VerdoesKleijn1999,vanBemmel2012}.\\

NGC 5141 was identified with the radio galaxy 4C +36.24 by \cite{caswell1967} and classified as a twin-jet Fanaroff-Riley type I (FRI) by \cite{owen1989}. It is also part of the FRI catalog (FRIcat; \citealt{FRIcat}), which was built by cross-correlating the seventh data release of the SDSS with the NRAO (National Radio Astronomy Observatory) VLA (Very Large Array) Sky Survey (NVSS), and the Faint Images of the Radio Sky at Twenty centimeters (FIRST).
Its radio emission, which is totally confined within the host galaxy, has an extension of only $\approx19$ kpc \citep{Emonts2010}.
The total flux density at 1.4 GHz is $S=880\pm30$ mJy \citep{Condon1998}, which corresponds to a luminosity\footnote{Calculated as follows: $L_{1.4}=10^{-23}4\pi D_L^2F_{obs}^{[Jy]}\nu_{1.4}/(1+z)^{\alpha+1}$ with $\alpha=-0.75$ measured by \cite{parma1999}.} of $L(1.4~GHz)=(8.2\pm0.3)\times10^{39}$ erg s$^{-1}$. 
4C +36.24 is a radiatively young source, as shown by \cite{parma1999}: in the 1.4-5 GHz spectral range, the source did not reveal any significant spectral slope change between the core and lobes, with a maximum spectral index value of $\alpha_{max}=-0.75$. They estimated a lower limit on the spectral break frequency ($\nu_{br}>30$ GHz), using the synchrotron-loss spectrum for the Jaffe \& Perola model \citep{jp}. Thus, they provided a radiative age upper limit of $t_{age}<4$ Myrs. 
At milliarcsecond resolution, \cite{giovannini2001} observed a one-sided jet in the same direction of the main large-scale jet and estimated an upper limit on the jet inclination angle of $\theta\leq58\degree$.

In the X-ray band, a single \textit{Chandra} observation of this source (November 25, 2003, ObsID 4055) is available. It was analyzed by \cite{Kharb}, who focused on NGC 5141 nuclear spectral analysis and found slightly extended emission ($\approx4"$) associated with the southern radio jet.
In this work, we analyze the same X-ray observation. Our main goal is to study  extended emission from NGC 5141 and the jet-ISM interaction.

\section{X-ray data analysis}
\label{sect_xray}
\begin{figure*}[!h]
 \centering
 \includegraphics[width=9.54cm] {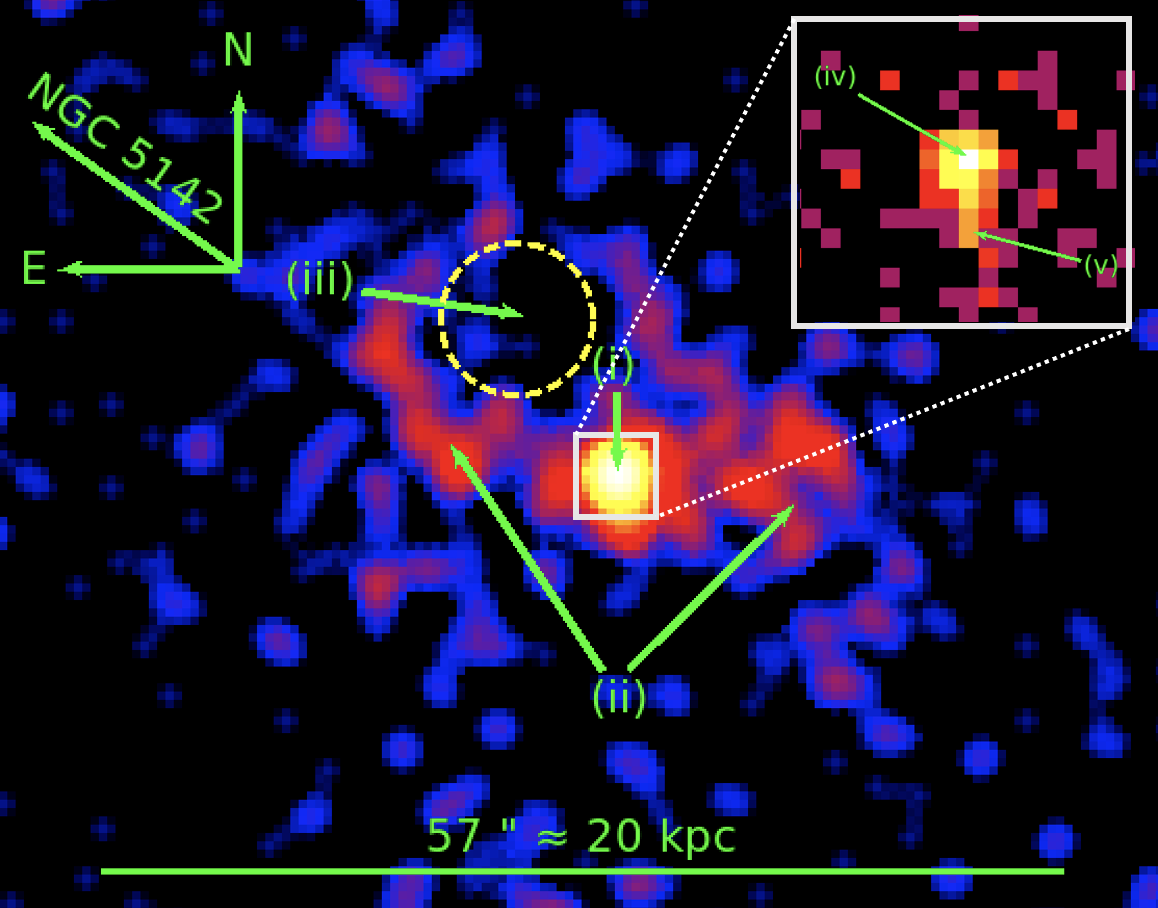}\quad
\includegraphics[width=8.2cm] {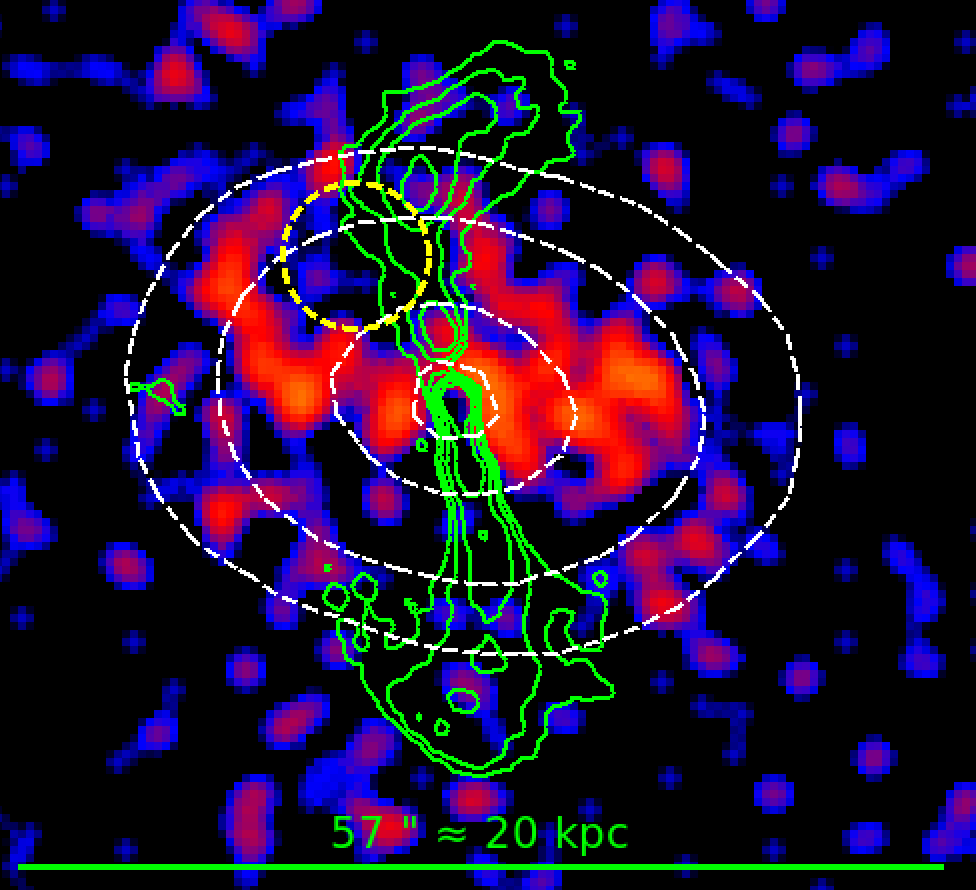}
\caption{NGC 5141 X-ray image in the 0.5-7.0 keV band  using a logarithmic scale with  $\sigma=$2 pixels of Gaussian smoothing and a scale parameter value $\geq$ 0.035  counts per pixel. \textit{Left panel:} Three main X-ray features can be identified: i) the nuclear region; ii) an extended diffuse gas emission (cavity rims and western blob); and iii) a cavity (dashed yellow circle). A zoom of the nonsmoothed nuclear region is shown in the top-right white panel: the core (iv) and the southern jet (v) are clearly visible. The green arrow points to the NGC 5142 optical center. \textit{Right panel:} Optical (dashed white) and radio VLA at 1.4 GHz (green; resolution$=$1.4", RMS$=$0.1 mJy/beam) contours overlap with the X-ray image, after having removed the nuclear emission (2"$\times$4" ellipse, see Section \ref{sect-extend-emis}). The cavity position is indicated  by the dashed yellow circle. The optical contours are taken from SDSS g-band emission within 70$\%$ of the NGC 5141 emitted light.  The radio green contours are at $-9, 9, 18, 36, 72, 144$ $\times$ RMS; the image is taken from the NRAO Science Data Archive.}\label{img-x-feat}
 \end{figure*} 

We analyzed the 31 ksec \textit{Chandra} ACIS-S observation of NGC 5141 available in the public archive. 
\textit{Chandra} data were reprocessed following standard procedures using the software {\sc CIAO}~4.11 with CALDB~4.8.1.  Good time intervals were defined by using the light curve to filter out times of background flares from the event file. The net exposure time was 26.8 ksec. In the 0.5-7 keV image, the most interesting feature is the northern X-ray cavity (Figure ~\ref{img-x-feat}-\textit{left panel}). If VLA radio contours at 1.4~GHz (taken from the NRAO public archive\footnote{\url{https://archive.nrao.edu/archive/archiveimage.html}}) are superimposed on the smoothed X-ray image, the X-ray cavity position corresponds to the northern radio lobe (see Figure ~\ref{img-x-feat}-\textit{right panel}).
In the southern hemisphere,  X-ray emission is generally lacking on the galactic scale. Only the nucleus is slightly elongated  to the south because of the southern jet emission.\\

For all the spectral analysis, the background was selected as an annular region with inner and outer radii of 50" and 70", respectively, within the same CCD  as the source but far away from the NGC 5141 X-ray structures (see Figure \ref{fig_cavity}). Point-like sources falling into the background region were removed.
The X-ray spectral analysis was performed with \textsc{xspec}12.11.1 \citep{Arnaud1996} in the 0.5-7 keV energy band.
The data were rebinned to have at least two counts per bin, and Poissonian statistics (Cash statistics - Cstat, \citealt{Cash1979}) was applied.
All the reported errors are at $1\sigma$ confidence level.

\subsection{Nuclear region: core and jet}

We analyzed the nucleus of the source. The X-ray spectrum was extracted from a circular region of $1.3\arcsec$ to exclude the abovementioned southern jet contribution. The nucleus best-fit model is a power-law that is absorbed by an intrinsic column density in addition to the Galactic one\footnote{NH$_{Gal}=1.02\times 10^{20}$ cm$^{-2}$, \cite{Kalberla2005}.} (Cstat=75.4, for 79 degrees of freedom, hereafter dof). Although the  total number of counts was limited (185), we could estimate the spectral index and the intrinsic absorption, $\Gamma=1.3\pm0.3$ and NH$_{int}=6\pm2\times10^{21}$ cm$^{-2}$, which are compatible with  the analysis of \cite{Kharb}. The intrinsic soft X-ray attenuation can be ascribed to the central dust lane detected with HST \citep{VerdoesKleijn1999,vanBemmel2012}.\\
The rest frame (unabsorbed) luminosity between 2-10 keV is $L_{2-10~keV}=(8_{-2}^{+4})\times10^{40}$ erg s$^{-1}$. Given the BH mass (see Section \ref{sec_source}), the Eddigton normalized X-ray luminosity is $L_{2-10 ~keV}/L_{Edd}=(2\pm1)\times10^{-6}$, where $L_{Edd}$ is defined as $1.26\times10^{38}$ M$_{BH}$/M$_{\odot}$ erg s$^{-1}$. This value is well compatible with inefficiently accreting sources \citep[e.g.,][]{Macconi2020}, in agreement with the LERG optical classification.

The jet spectrum was extracted within a box (2.5" $\times$ 3") located at 1.3$\arcsec$ from the nucleus. 
The spectrum consists of 27 counts. The data were fitted with a power-law of spectral index $\Gamma_{jet}=2.2\pm0.4$ and a Galactic absorber (Cstat=10.9 for 10 dof).
The jet luminosity between 2-10 keV is $L_{jet}=(2.4^{+0.6}_{-0.5})\times 10^{39}$ erg s$^{-1}$, which is one order of magnitude lower than the nuclear one, as is generally observed in X-ray jets \citep[see e.g.,][]{Sambruna2006}.

\subsection{Extended emission}
\label{sect-extend-emis}

  \begin{figure}[!h]
 \centering
 \includegraphics[width=6cm] {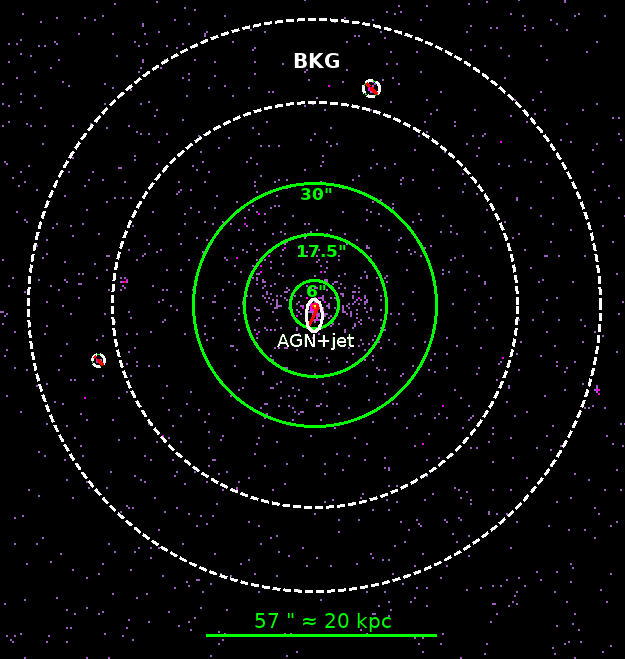}
 \caption{0.5-7 keV image of NGC 5141 in logarithmic scale: the green annuli represent the extraction regions used to perform the spatially resolved spectral analysis of the diffuse emission. The dashed white annulus is the chosen background region.}
 \label{fig_cavity}
 \end{figure}

\begin{table*}[!h]
\footnotesize
\renewcommand{\arraystretch}{1.4}
\centering
\begin{tabular}{lcrrrrr}
\hline
\multicolumn{1}{c}{r$_{m}-$r$_{M}$} 
&\multicolumn{1}{c}{Net Counts} 
&\multicolumn{1}{c}{kT} & 
\multicolumn{1}{c}{N$_{\rm APEC}$} & 
\multicolumn{1}{c}{$\Gamma$} & 
\multicolumn{1}{c}{N($\Gamma$)} & 
\multicolumn{1}{l}{Cstat/dof}\\
\multicolumn{1}{c}{[arcsec]}&\multicolumn{1}{c}{[$0.5-7$ keV]}&\multicolumn{1}{c}{[keV]}&&&\multicolumn{1}{c}{{\tiny[phot/keV/cm$^2$/s] @1 keV}}\\
\multicolumn{7}{c}{\bf{ Projected spectra:} \textsc{\tiny TBABS*(POWER-LAW+APEC)}}\\
\hline
0-30 & 429& $0.80\pm0.05$& $(8\pm1)\times10^{-6}$ &$1.5\pm0.3$& $(6\pm2)\times10^{-6}$ & 68.8/108\\
\multicolumn{7}{c}{\bf{ Deprojected spectra:} \textsc{\tiny PROJCT*TBABS*APEC+TBABS*POWER-LAW}}\\
0-6 & 48 & $0.80$F & $<3\times10^{-7}$ & $1.5$F & $(1.6\pm0.4)\times10^{-6}$ & 15.7/20\\
6-17.5 & 210 & $0.80$F& (3.8$\pm0.8$)$\times10^{-6}$ & $1.5$F & $(3.6\pm0.6)\times10^{-7}$ &47.1/70 \\
17.5-30 & 167 & $0.80$F & $(4.5\pm0.9$)$\times10^{-6}$& $1.5$F & $(1.2\pm0.7$)$\times10^{-7}$ & 43.2/59\\ 
&&&&&total fit statistic& 117.7/149\\
\hline
\end{tabular}
\normalsize
\caption{Results of the extended emission spectral analysis as described in Section \ref{sect-extend-emis}. The reported columns represent: region size for each extracted spectrum (minimum and maximum radii of the circular annuli), counts number, temperature of the \textsc{apec} model, \textsc{apec} normalization, spectral index of the \textsc{power-law} model, \textsc{power-law} normalization, statistics of the fit.}
\label{tab-x-ray-analysis}
\end{table*}

A spectrum was extracted in a circular region of 30 $\arcsec$ centered on the nucleus, after having excluded an elliptical region (semi-minor axis=2$\arcsec$, semi-major axis= 4$\arcsec$) containing nucleus and jet (see Figure \ref{fig_cavity}). The selected region fully contains the galactic X-ray emission and roughly corresponds to the optical galactic scale (see Figure \ref{img-x-feat}-\textit{right panel}) \\
The diffuse emission was fitted with a thermal  model (\textsc{apec}, Astrophysical Plasma Emission Code, \citealt{Smith2001}), attenuated by Galactic absorption. 
Given the poor statistics, we could not constrain the metal abundance that was kept fixed to the solar value.
Although the model was acceptable (Cstat=106.6 for 110 dof), an excess of photons above 1 keV was still present in residuals. A power-law was then added to the thermal model. The fit strongly improved, giving a C-stat value of 68.8 for 108 dof. 
The best-fit parameters were kT=$0.80\pm0.05$ keV for the diffuse emission and $\Gamma=1.5\pm0.3$ for the power-law. The hard slope and the 0.5-8 keV power-law luminosity (calculated in the energy range usually adopted in literature), $L_{0.5-8~keV}=(2.8\pm0.8)\times10^{40}$ erg s$^{-1}$, are fully consistent with unresolved emission from bright low-mass X-ray binaries (LMXBs) typically observed in early-type galaxies \citep{Irwin2003,Boroson,Lehmer2019}.\\

As we were interested in the properties of the gas surrounding the cavity, our best-fit model (\textsc{apec}+\textsc{power-law}) was applied individually to three spectra extracted from concentric annuli with  6, 17.5, and 30 arcseconds radii, respectively (see Figure  \ref{fig_cavity}).  Low counts preclude fitting several parameters to each spectrum, so we kept $\Gamma$ fixed at 1.5 while fitting to find the temperature in each annulus. We obtained: $kT_{0-6"}=0.6_{-0.3}^{+0.2}$ keV, $kT_{6-17.5"}=0.89\pm0.07$ keV, and $kT_{17.5-30"}=0.75_{-0.09}^{+0.08}$ keV. Thus, since in all three annuli the temperatures are compatible with that in the 0-30" region, in each annulus both the temperature (kT=0.80 keV) and power-law index ($\Gamma=1.5$) were frozen.

In order to take projection effects into account, the multiplicative model \textsc{projct} was applied to the \textsc{apec} component (see Table \ref{tab-x-ray-analysis} for model description and best-fit values). The \textsc{projct} model deprojects 2D spectra, assuming spherical symmetry.
The deprojected \textsc{apec} normalization (N$_{\rm APEC}$) is directly connected to the thermal component emission measure\footnote{The Apec normalization is N$_{\rm APEC}$=$\frac{10^{-14}\int n_e n_H dV}{4\pi [D_A(1+z)]^2}$ with $D_A=\frac{D_L}{(1+z)^2}$ being the angular diameter distance (cm), $D_L$ the luminosity distance ($D_L=75.5$ Mpc with the adopted cosmology), $z$ the redshift of the source, V the volume of the emitting region, and n$_e$ and $n_H$ the electron and H densities (cm$^{-3}$), respectively. The number densities ($n_e$, $n_H$) are integrated over the spherical shell volume.}: $EM=\int n_en_H dV$. 
For a collisional ionized plasma with solar abundance, $EM=0.82n_e^2\times V$ \citep[see e.g.,][]{gitti2012} \footnote{ the total particles density $n=n_e+n_p$ (i.e., neglecting nuclei)} is calculated through the formula $n=(2+1/2\frac{Y}{X}+1/2\frac{Z}{X})n_p$, where X, Y, and Z are the mass fraction of hydrogen, helium, and metals, respectively, thus, given  N$_{\rm APEC}$, the gas density
 \begin{equation}
 n_e=\sqrt{\frac{4\pi \cdot 10^{14}\cdot N_{\rm APEC} \cdot [D_L/(1+z)]^2 }{0.82 \cdot V}} \label{paso}
\end{equation}
can be obtained for each shell. The n$_e$ radial profile is shown in the {\it left panel} of Figure  \ref{img_radial_profiles}. 
The power-law was left out of deprojection with its slope frozen to 1.5 and its normalization free to vary \citep[see][for an analogous analysis procedure]{su2017,Lakhchaura2018}. Spectral results are reported in Table \ref{tab-x-ray-analysis}. \\
We also tested a deprojected model for both the power-law and thermal components (\textsc{projct*tbabs*(power-law+apec)}), which is valid in  the case that  the LMXBs have  a spherical distribution. Also in this case, we froze both kT and $\Gamma$ values to the 0-30" ones. The deprojected \textsc{apec} normalization is fully compatible with our best-fit model (see Table \ref{tab-x-ray-analysis} for comparison): $N_{0-6"}<3\times10^{-7}$, $N_{6-17.5"}=3.8_{-0.9}^{+0.8}\times10^{-6}$ and $N_{17.5-30"}=4.5_{-0.9}^{+1.0}\times10^{-6}$.

\subsection{X-ray cavity significance}
\label{sec-significance}
The most intriguing feature in the X-ray image is the depression partially coincident with the northern radio lobe, suggestive of an interaction between the lobe plasma and the surrounding gas (see Figure \ref{img-x-feat}).
It cannot be excluded that an X-ray cavity overlapping with the southern lobe is also present. However,  the low surface brightness and the lack of bright surrounding hot gas make the detection difficult.  A deeper X-ray observation is necessary to confirm whether the jet has excavated the ISM around the southern radio lobe.\\

We assumed that the northern X-ray cavity is spherical with a radius of  4.5 $\arcsec$ (yellow dashed circle in Figure \ref{img-x-feat}). To test the significance of the depression, we compared the  number of counts per unit area inside the cavity (0.094 counts arcsec$^{-2}$) with a control region.  The control region is chosen as  the annulus extending from 6" to 15" from the AGN (i.e., spanning the radial extent of the cavity region), from which the cavity region and its surrounding bright rims were excluded. The resulting control region surface brightness is 0.255 counts arcsec$^{-2}$ (the control region, i.e., the annulus minus rims, is shown in Figure \ref{stat_reg} in Appendix \ref{significance}). Therefore, the surface brightness is found to  be reduced by $\approx60\%$ within the X-ray cavity with an associated significance level of $\approx4 \sigma$ \citep[following the method in][]{Ubertosi2021}\footnote{The percentage depression (D) was calculated as follows: $D=(1- \frac{S_{CAV}}{S_{CR}})\times100$, where $S_{CAV}$ is the cavity surface brightness and $S_{CR}$ is the control region surface brightness. The cavity significance was calculated as $D/\Delta D$, where $\Delta D$ was calculated through surface brightness propagation of uncertainty.}. Compared to the typical drops observed in clusters (i.e., $\approx20-40 \%$ below the surrounding gas level), the brightness depression is higher. The strong anisotropic NGC 5141 gas distribution could lead to the lack of galactic gas in the cavity background-foreground regions (if compared to sources with richer gas amount, e.g., clusters) and it could be responsible for such a deep depression. \\ However, we also performed additional tests to  corroborate our result, exploiting different control regions. As detailed in Appendix \ref{significance}, we found a depression percentage varying between 40-70$\%$ and a significance ranging between 2-5$\sigma$.

These tests strengthen the X-ray cavity detection. Considering the low statistics and the less uniform gas distribution in this galaxy than in clusters, we are confident that the cavity in this galaxy is reliable (see also Section \ref{sec_literature}).

\section{AGN heating versus cooling}
\subsection{Heating}
\label{heating}
\begin{figure}
\centering
\includegraphics[width=4.5cm] {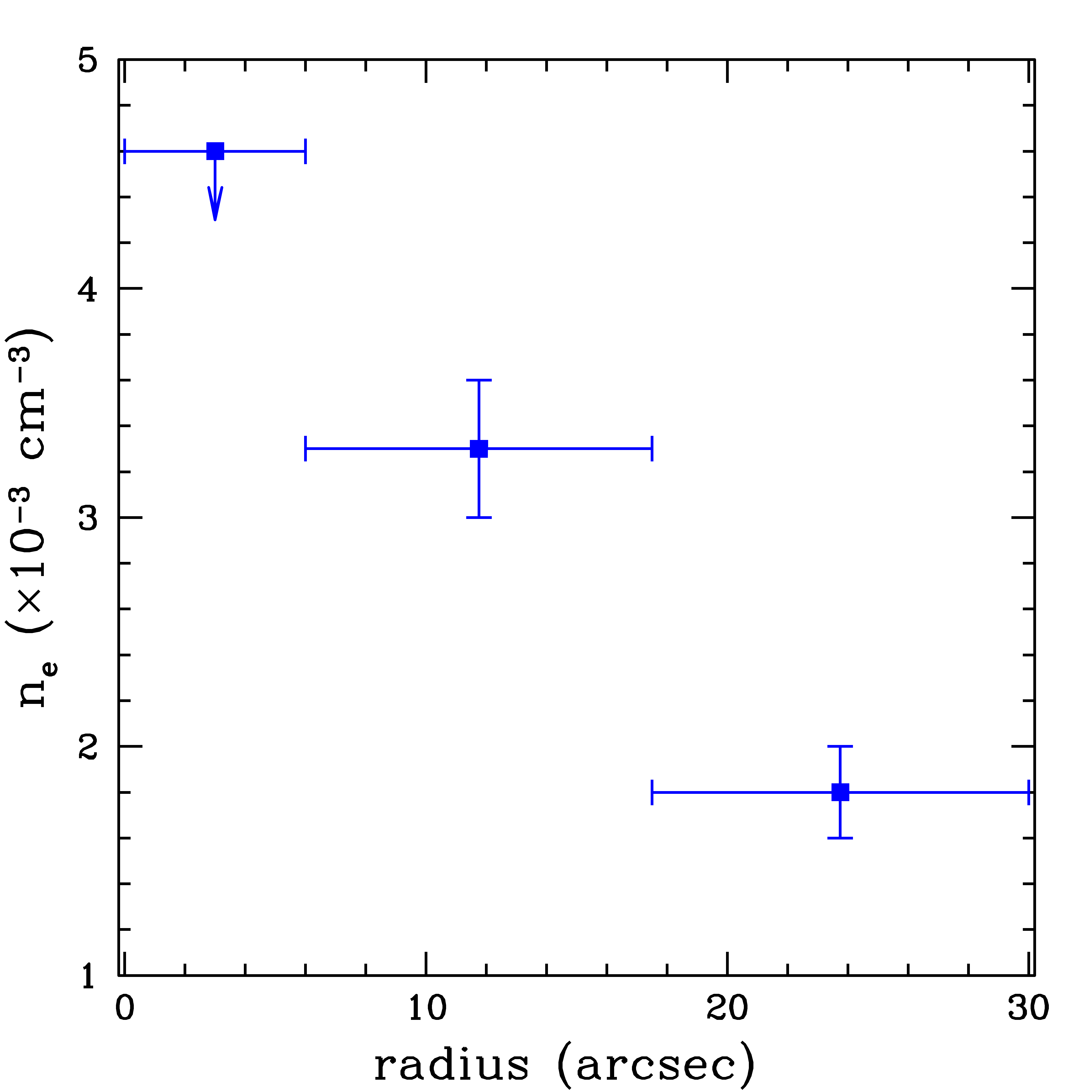}\qquad \hspace{-0.8cm}
\includegraphics[width=4.5cm] {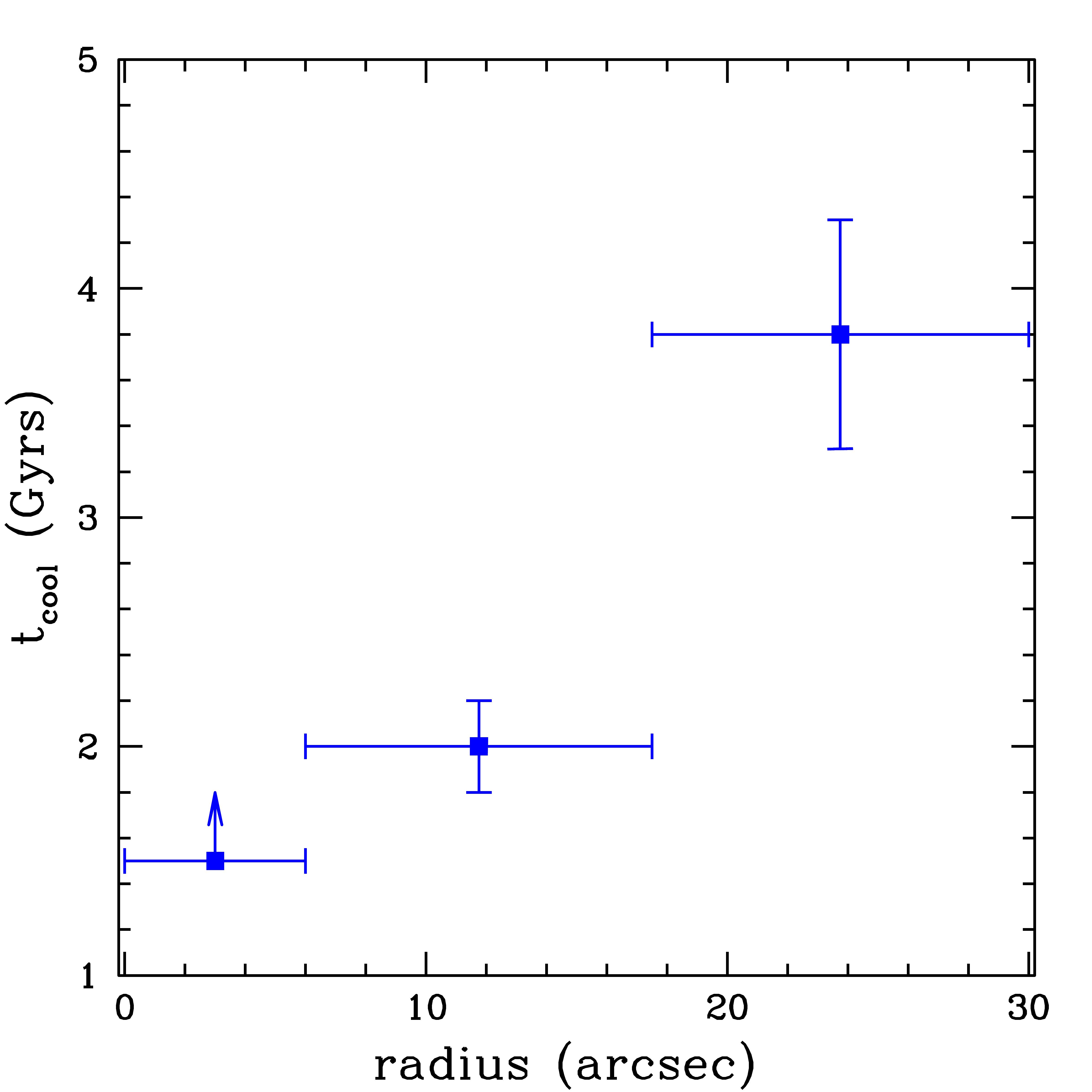}
\caption{Radial profile of density (\textit{left panel}) and cooling time (\textit{right panel}) of the three concentric regions as calculated in Section \ref{sec_cooling} and described in Table \ref{tab-x-ray-analysis}.}\label{img_radial_profiles}
\end{figure}

A study of jet-ISM feedback needs the gas surrounding the cavity to be characterized.
As shown in Figure  \ref{img_radial_profiles} ({\it left panel}), the ISM density decreases with increasing distance from the center. In the 6"-17.5" annulus, where the cavity is found, the density is $n_e=(3.3\pm0.3)\times 10^{-3}$ cm$^{-3}$. Given $n_e$, and adopting n=n$_p$+n$_e\sim 1.82n_e$ \citep[e.g.,][]{gitti2012}, the gas pressure can be calculated as  $P=nkT\approx1.82n_ekT\approx(7.7\pm0.9)\times10^{-12}$ erg cm$^{-3}$.\\
The energy required to excavate the medium is the sum of the work done by the lobes (PV) and their thermal energy \citep{mcnamara2007}. It can be expressed as $$E_{cav}=\frac{\gamma}{\gamma -1}PV_{cav},$$ with $\gamma$ being the ratio of the specific heats and V$_{cav}$ being the volume of the cavity.
For $\gamma=4/3$, which is appropriate for a relativistic plasma, and a spherical cavity of radius $R_{C}=4.5\arcsec$ (see the dashed yellow circle in Figure \ref{img-x-feat}), we obtain: $E_{cav}=(1.5\pm0.2)\times10^{55}$ erg.\\

To determine the average jet power required to push  aside the ISM gas, an estimate of the cavity age is necessary.
We considered three different methods \citep{birzan2004} to calculate the time (t): i) the cavity is assumed to rise  through the hot gas atmosphere at the sound speed ($t_s$); ii) the cavity is assumed to rise buoyantly ($t_b$); and iii) the cavity age is the time required to refill the displaced volume ($t_r$).\\
In detail, $t_{s}$ is expressed as the ratio between the projected distance of the cavity from the center (D=10.5") and the sound speed $c_s=\sqrt{\gamma kT/\mu m_p}$, with $\gamma=5/3$ (monoatomic gas) and $\mu$, the mean molecular weight, being equal to 0.61 once that the solar abundance is assumed.
In the second case, the estimated age is expressed as $t_b=D/\sqrt{\frac{2gV_{cav}}{SC}}$,  where the gravitational acceleration is estimated as g=$2\frac{\sigma_{\star}^2}{D}$  \citep{birzan2004}. Here the stellar velocity is $\sigma_{\star}=242\pm4$ km/s as provided by Data Release 16 (DR16) of the SDSS\footnote{$https://www.sdss.org/dr16/$}. The term $C$ is the drag coefficient assumed to be 0.75 \citep{churazov2001}, and $S=\pi R_{C}^2$ is the X-ray cavity cross section.
The refill time was calculated as $t_r=2\sqrt{R_{C}/g}$.
The estimated ages and the resulting cavity power, $P_{cav}=E_{cav}/t_{cav}$, are listed in Table \ref{tab-p-cav}.

 We note that there are systematic errors for two of the age estimates. Projection effects mean that the distance from the AGN to the center of the cavity is likely underestimated, leading to underestimates of the ages. As reported by \cite{giovannini2001}, the jet inclination angle of 4C +36.24 should be 58$\degree$ or less. On the other hand, we have assumed that the radio bubble was formed close to the AGN and traveled to its current location.  If the bubble was excavated by the jet further from the core, these ages may be overestimates.

\begin{table*}[!h]
\setlength{\arrayrulewidth}{0.2mm}
\setlength{\tabcolsep}{4pt}
\renewcommand{\arraystretch}{1.0}
\begin{center}
\begin{tabular}{p{1cm}p{2cm}p{3cm}}
\hline
&\multicolumn{1}{c}{Age} & $P_{cav}$\\
&\multicolumn{1}{c}{[Myrs]} & [erg s$^{-1}$]\\
\hline
$t_s$&\multicolumn{1}{c}{$7.9\pm0.3$} &$(6.0\pm0.8)\times10^{40}$\\ 
$t_b$&\multicolumn{1}{c}{$8.6\pm0.3$} & $(5.6\pm0.8)\times10^{40}$\\ 
$t_r$&\multicolumn{1}{c}{$14.0\pm0.3$}&$(3.4\pm0.5)\times10^{40}$\\\hline\\\hline
 \end{tabular}
\normalsize
\caption{Cavity power calculated considering the three different age methods (sound, buoyant and refill times, see text for details). The cavitiy energy is E=$(1.5\pm0.2)\times10^{55}$~erg.} \label{tab-p-cav}.
\end{center}
\end{table*}

Since different metallicities can alter the \textsc{apec} normalization (N$_{\rm APEC}$) and the X-ray fit cannot constrain the gas metallicity, we considered various metallicity values for the hot gas thermal emission in the fit and we recalculated the cavity power. 
Even considering an extensive range of possible metallicities, from 0.3 to 2 solar abundances, $P_{cav}$ is found to range from $\approx9.6\times10^{40}$ erg s$^{-1}$ to  $\approx3.5\times10^{40}$ erg s$^{-1}$, respectively. Thus, the estimated cavity power is within a factor 2 from our best-fit estimate (solar abundance): given the large uncertainties in the $P_{cav}$ estimate, the metallicity does not have a tremendous impact.\\

The cavity age ($t_{cav}\sim 8-14$ Myrs) is larger than the upper limit on the radiative age $t_{rad}<4$ Myrs, found by \cite{parma1999}. We notice that the radiative age would be compatible with the cavity sound expansion time ($t_{se}$), that is the time that is necessary for the cavity to reach the current size, expanding at the sound speed, and assuming that it was produced in the current position: $t_{se}=R_{C}/c_s\approx3.3$ Myrs. 

The source radiative age was estimated by using the radio lobes spectral index steepening between 1.4 and 5 GHz (interpreted as electron aging due to synchrotron and inverse Compton losses) and by calculating a magnetic field strength through the "minimum energy assumption". The energy break was estimated as a lower limit (> 30 GHz) for this source, implying that only an upper limit on the radiative time can be obtained. Due to the discrepancy between $t_{rad}$ and $t_{cavity}$, a comparison between X-ray cavity power and radio power is prevented. 

\subsection{Cooling}
\label{sec_cooling}
\begin{figure}[!h]
 \centering
 \includegraphics[width=9cm] {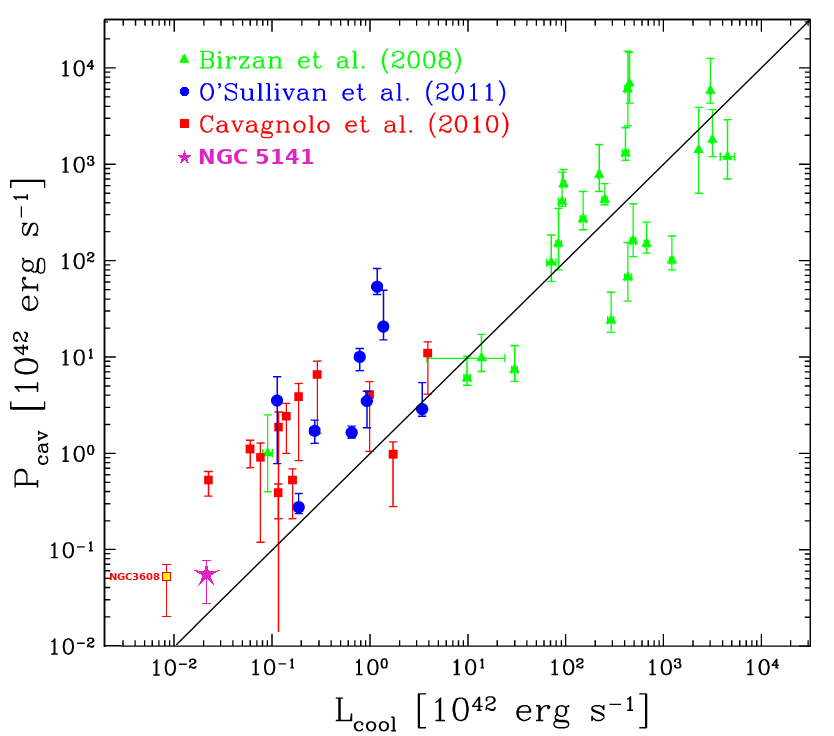}
 \caption{X-ray cooling luminosity (see Section \ref{sect_xray}) versus cavity power (calculated as $P_{cav}=E/t_b$) from \cite{gitti2012}. NGC 5141 is the magenta star (see Table \ref{tab-p-cav} for values). The cavity energy is estimated as $E=4pV$. Different symbols denote systems in different samples: green triangles are clusters from \citealt{Birzan2008}, red squares are giant ellipticals from \citealt{Cavagnolo2010} and blue circles are groups from \citealt{o'sullivan2011}. NGC 3608, which is part of the \citealt{Cavagnolo2010} sample, is highlighted as the yellow-filled square with red edges. The diagonal line indicates $P_{cav}=L_{cool}$. If cooling is exactly balancing heating, the content of the X-ray cavities is fully relativistic and $P_{cav}=P_{jet}$, thus the data points should be on the line.}\label{fig_nulsen7}
 \end{figure}

Once the ISM temperature and density values are known, the time required by the ISM to efficiently cool through radiative emission can be calculated for each annulus as follows: 
\begin{equation}
 t_{cool}=\frac{\gamma}{\gamma-1}\frac{k~T}{\mu X n_e \Lambda(T,Z)} 
\end{equation}
Here $\gamma=5/3$, $\mu\approx0.61$, and X is the hydrogen mass fraction ($X=0.71$ for solar abundances).
According to the source temperature and metallicity, the cooling function is $\Lambda=3.5\times10^{-23}$ erg~cm$^3$~s$^{-1}$ \citep{Sutherland1993}. The estimated cooling times are shown in Figure \ref{img_radial_profiles} ({\it right panel}).
The cooling radius ($r_{cool}$) is generally defined as the radius within which the cooling time equals the age of the system ($t_{sys}$).
It delimits the region where the ISM effectively cools and where the cooling luminosity ($L_{cool}$), that is the thermal X-ray bolometric luminosity (0.1 - 100 keV), is measured. In order to compare our results with similar systems found in literature, we chose the commonly adopted 7.7 Gyrs as the maximum cooling time of the system (corresponding to the lookback time at z=1, see, e.g., \citealt{birzan2004,Birzan2008,Cavagnolo2010,Eckert2021}). Since this is longer than the cooling time of each analyzed annulus, the entire 0-30" region lies inside the cooling radius; thus, we could estimate the ISM cooling luminosity from the corresponding spectra (excluding the AGN and jet), finding: $L_{cool}=(2.1\pm0.3)\times10^{40}$ erg s$^{-1}$. Conversely, if we consider 3 Gyrs as the threshold (also proposed as a typical cooling time threshold within groups and clusters hosting cavities, see \citealt{Panagoulia2014}), the cooling luminosity would be calculated within the two innermost regions (within 17.5", see Figure  \ref{img_radial_profiles}-{\it right panel}). Its value would not have changed dramatically (within a factor 2): $L_{cool}\approx1.2\times10^{40}$ erg s$^{-1}$. \\
Given both NGC 5141's X-ray cavity power and its cooling luminosity, it is possible to investigate whether the system is balanced or not. Since the two quantities are comparable (i.e., $P_{cav}/L_{cool}=3\pm1$), we conclude that the heating power is able to balance the cooling within NGC 5141.   
In Figure \ref{fig_nulsen7}, taken from \cite{gitti2012}, the X-ray cooling luminosity versus the cavity power is shown for different systems with cavities (clusters from \citealt{Birzan2008}, groups from \citealt{o'sullivan2011} and giant elliptical galaxies from \citealt{Cavagnolo2010}).
NGC 5141 (magenta star), included in the plot, follows the general trend. 
\label{similar-objects}
 \begin{figure*}[!h]
 \centering
 \includegraphics[width=6.1cm] {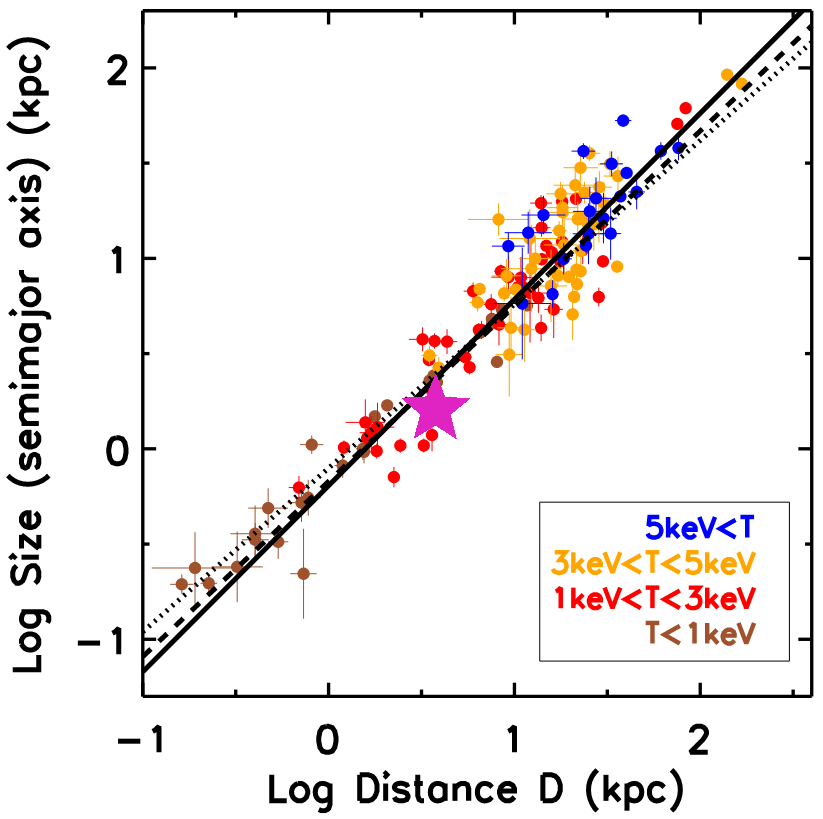}\qquad\hspace{-0.75cm}
 \includegraphics[width=6.1cm] {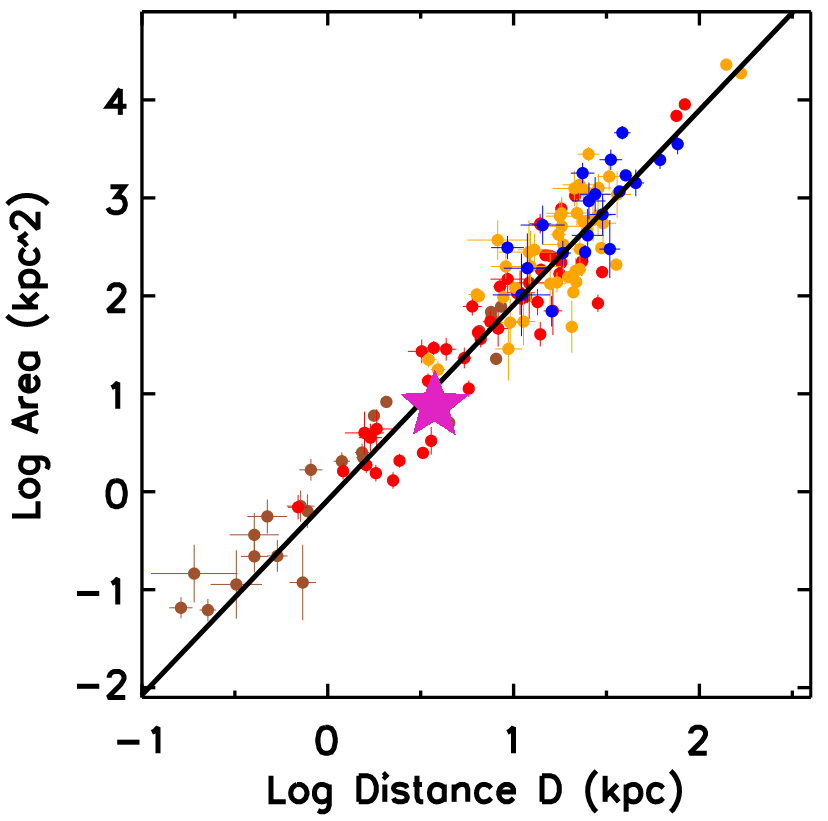}\qquad\hspace{-0.75cm}
  \includegraphics[width=6.1cm] {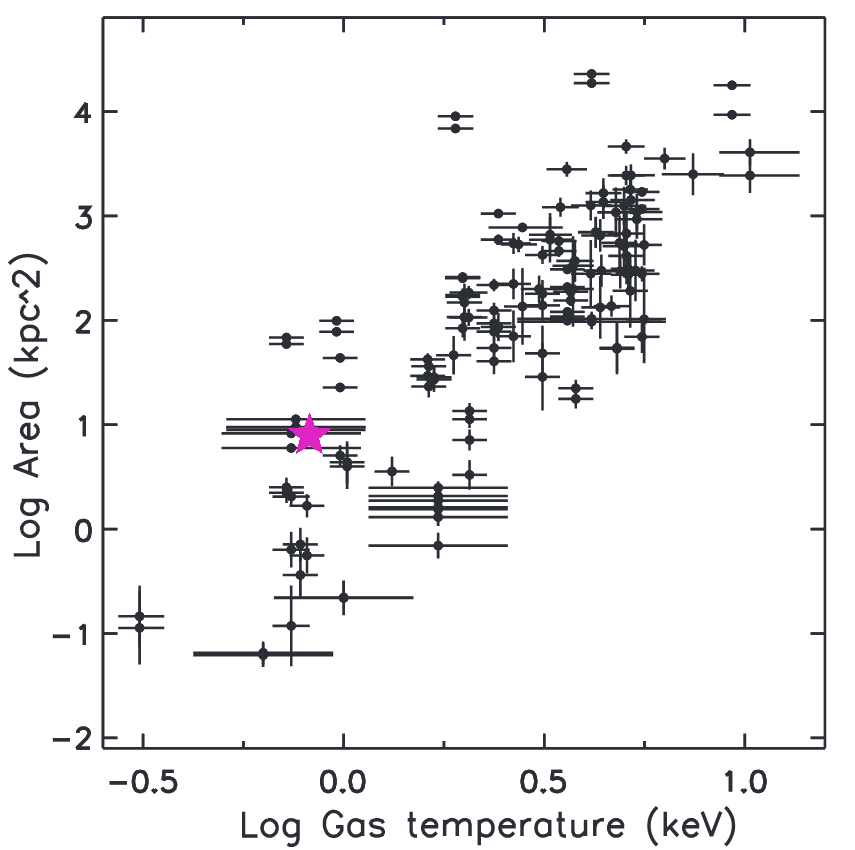}
  \caption{Plots adapted from \citealt{shin2016}, which show their large X-ray cavity sample properties, as observed in various environments (isolated galaxies, groups, and clusters). The NGC 5141 (the magenta star) size, distance from the center, and temperature are compared with this sample. {\it Left panel:}  Distance from the center versus linear size for cavities. {\it central panel:}  Distance versus area for cavities. {\it right panel:} X-ray gas temperature versus  cavity area.}\label{shin1}
 \end{figure*}
 \subsection{Comparison with the literature}
 \label{sec_literature}
\cite{shin2016} collected a large sample of cavities (148 in total), detected by \textit{Chandra} with exposure times ranging from a few to hundreds of kilo-seconds. Their sample covers an extensive dynamical range (from isolated galaxies to rich clusters). NGC 5141 can be included in their relations: in Figure  \ref{shin1}, the cavities' distance from the system center is compared to their linear size ({\it left panel}) and area ({\it central panel}), while in the {\it right panel} the system gas temperature is plotted against the cavity area. NGC 5141 is indicated through a magenta star. It follows all the correlations found by \cite{shin2016} with a good agreement:  this indirectly supports the reliability of the cavity detection in NGC 5141. \\

To the best of our knowledge, there are only two galaxies in  the literature characterized by  as low  a total cavity power as NGC 5141 (i.e., $<10^{41}$ erg s$^{-1}$): NGC 3608 (part of the \citealt{Cavagnolo2010} sample, see Figure \ref{fig_nulsen7}) and NGC 4477.
NGC 3608 is a galaxy showing cavities with similar $P_{cav}$  to NGC 5141 ($\approx10^{40}$ erg s$^{-1}$ in total, \citealt{Cavagnolo2010}; it is  represented in Figure \ref{fig_nulsen7}  by the yellow-filled red square in the bottom left corner). NGC 4477 hosts the smallest cavity structures known in terms of size, distance from the center, and cavity power ($\approx10^{39}$ erg s$^{-1}$, see \citealt{Li2018}).  As mentioned in Section \ref{introduction}, it along with NGC 193 are the only two S0 galaxies with cavities known (NGC 193 shows three prominent cavities with a high cavity power: $\approx10^{42}-10^{43}$, \citealt{bogdan2014}). 
NGC 4477 is not in Figure \ref{fig_nulsen7}: a direct comparison with the other systems is not possible since a different cooling time threshold was adopted \citep{Li2018}. However, there are important differences with NGC 5141: (i) both NGC 3608 and NGC 4477 have a cooler ISM (kT$\approx0.3$ keV); (ii) they show multiple cavities within their galaxy; and, most importantly, (iii) {\it they do not exhibit extended radio emission, precluding direct evidence for a connection between jet activity and X-ray cavities}  and they are therefore named ghost cavities \citep{birzan2004}). NGC 3608 was not detected in  the radio band  recently, even with LOFAR \citep{birzan2020}.
In conclusion, NGC 5141 represents a unique example of  a galaxy possessing a low-power X-ray cavity clearly connected to jet activity. This enables us to expand the $P_{cav}/L_{cool}$ relation  at the lower limits in terms of power for radio-filled sources.

\section{Summary and Conclusions}
We report the discovery of an X-ray cavity within the ISM of NGC 5141. The X-ray cavity is located in the northeast region of the galaxy within a few kiloparsecs from the central AGN, and it partially corresponds to the northern radio lobe. NGC 5141 is classified in the radio band as a small low-power FRI whose emission is contained within the optical galaxy. \\
The X-ray cavity size, area, and distance from the AGN, as well as the temperature of the surrounding medium, follow the correlations found by \citealt{shin2016} (Figure \ref{shin1}). 
This agreement, together with the tests discussed in Section \ref{sec-significance} and Appendix \ref{significance}, corroborates the robustness of our cavity detection.\\
The total energy required to inflate the X-ray cavity is $E_{cav}=(1.5\pm0.2)\times10^{55}$ erg.  Assuming that it was produced in the core surroundings and that it rose buoyantly to the current position, a cavity age can be estimated as $t_b=8.6\pm0.3$ Myrs (see also Table 2 for the cavity ages estimated following different methods). Then, the cavity power is $P_{cav}=E_{cav}/t_b=(5.6\pm0.8)\times10^{40}$ erg s$^{-1}$. We note that $P_{cav}$ can be compared to the cooling luminosity of the system ($L_{cool}=(2.1\pm0.3)\times10^{40}$ erg s$^{-1}$, i.e., X-ray bolometric luminosity inside $r_{cool}$, see Section \ref{sec_cooling}) to evaluate the heating-cooling balance: in Figure \ref{fig_nulsen7}, the cavity power and the cooling luminosity of NGC 5141 are compared to those observed in clusters, groups, and galaxies. Our source, in the  lower left corner of the plot, follows the $P_{cav}=L_{cool}$ correlation, attesting  that this connection is present within systems ranging from $P_{cav}\sim 10^{46}$ erg s$^{-1}$ down to $10^{40}$ erg s$^{-1}$.\\

NGC 5141 represents a unique source because: (i) its single cavity  has the weakest $P_{cav}$ value among known radio-filled systems, allowing us to confirm the heating-cooling balance at the lowest edge of radio galaxy power and size; and (ii) it represents a rare signature of jet-ISM interaction found in a small lenticular-elliptical galaxy. If it is a lenticular, it would be the third S0 galaxy hosting an X-ray cavity after NGC 193 and NGC 4477.

\section*{Acknowledgements}
DM gratefully acknowledges the INAF-OAS Ph.D. fellowship.
The authors wish also to thank the anonymous referee for useful comments and suggestions that have
improved the paper.

\bibliographystyle{aa}
\bibliography{biblio}
\appendix
\section{X-ray cavity detection significance}
\label{significance}
\begin{figure}[!h]
 \centering
\includegraphics[width=8.8cm] {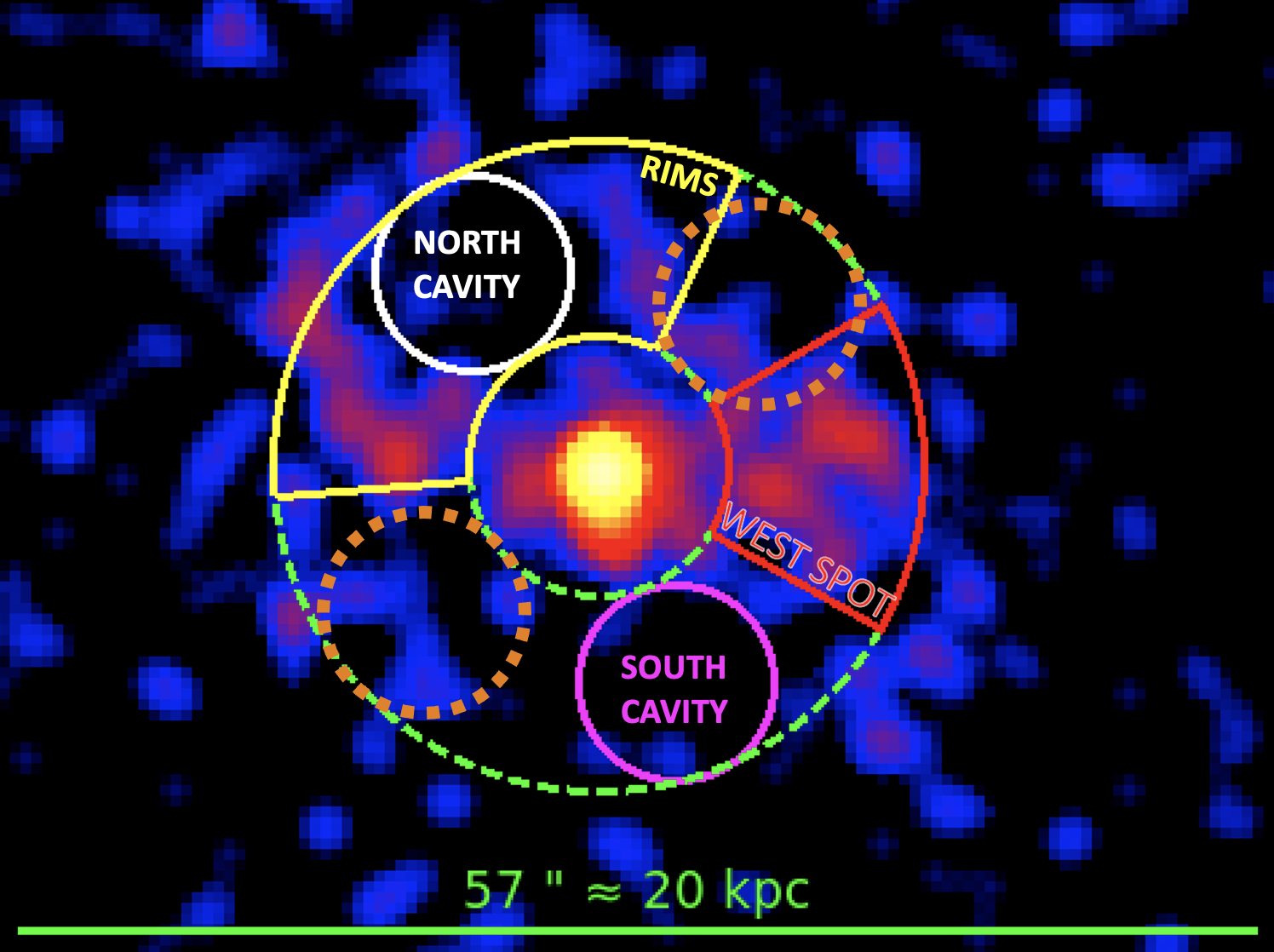}
\caption{NGC 5141 X-ray image in the 0.5-7.0 keV band in logarithmic scale with  $\sigma=$2 pixels of Gaussian smoothing and scale parameter value $\geq$ 0.035  counts per pixel. The different control regions are highlighted: the 6"-15" annulus is in dashed green, the north cavity region is in white, the rim region is in yellow, the west spot region is in red, a possible south cavity is in purple and the two outer circles are in dotted orange.}\label{stat_reg}
 \end{figure}

We evaluated the X-ray cavity depression compared to the surrounding emitting gas level and its relative significance, exploiting three different control regions in addition to the one reported in Section \ref{sec-significance} (i.e., annulus minus rims):
\begin{enumerate}
    \item  an annulus extending from 6" to 15" from the AGN (the same as in Section \ref{sec-significance}) from which we subtracted the cavity region itself (control region = annulus minus north cavity, see Figure \ref{stat_reg} where all the regions are highlighted). The resulting depression is $\approx67\%$ with a 4.9$\sigma$ significance;
    \item we removed all the brightest extended emission from the control region: we subtracted both the cavity rims and the bright west spot from the 6"-15" annulus region (control region = annulus minus rims minus west spot). In this case, the cavity depression is $\approx39\%$ with a significance of 1.5$\sigma$;
    \item from the same control region we excluded a putative southern cavity supposed to be at the same distance from the center as the northern cavity, with the same size and overlapping with the radio lobe emission (control region = annulus minus rims minus west spot minus south cavity). In this case, the depression is on the order of $\approx45\%$ with a significance of 1.9$\sigma$.
\end{enumerate}

Furthermore, the significance level of the X-ray cavity detection was also checked with two additional  circular regions, following \cite{Pasini2021}: the net number of counts in the cavity (six) was compared to those contained in two outer circles with the same cavity area (orange dotted circles in Figure \ref{stat_reg}). They were selected to be the nearest to the cavity region, at the same distance from the central AGN, but completely avoiding the emission from the bright rims that envelop the cavity. 
The significance of the cavity detection was calculated as a signal-to-noise ratio: $\frac{N_O-N_I}{\sqrt{N_O+N_I}}$, where $N_I$ is the number of counts inside the cavity, and $N_O$ is the counts number of the region outside the cavity. The procedure is the same for the two circular regions adjacent to the cavity. 
In this way, we obtained $\approx2\sigma$ significance comparing the cavity counts with both control regions.

\end{document}